# A Contextual Approach to Technological Understanding and Its Assessment

Eline de Jong[1*] & Sebastian De Haro[2]

## Abstract

Technological understanding is not a singular concept but varies depending on the context. Building on De Jong and De Haro's (2025) notion of technological understanding as the ability to realise an aim by using a technological artefact, this paper further refines the concept as an ability that varies by context and degree. We extend its original specification for a design context by introducing two additional contexts: operation and innovation. Each context represents a distinct way of realising an aim through technology, resulting in three types (specifications) of technological understanding. To further clarify the nature of technological understanding, we propose an assessment framework based on counterfactual reasoning. Each type of understanding is associated with the ability to answer a specific set of what-if questions, addressing changes in an artefact's structure, performance, or appropriateness. Explicitly distinguishing these different types helps to focus efforts to improve technological understanding, clarifies the epistemic requirements for different forms of engagement with technology, and promotes a pluralistic perspective on expertise.

**Funding:** This publication is an outcome of the project Quantum Impact on Societal Security, project number NWA.1436.20.002, which is funded by the Dutch Research Council (NWO), The quantum/nanorevolution.

[1,2] Institute for Logic, Language and Computation; Institute of Physics; Qusoft Research Center for Quantum Software; University of Amsterdam, The Netherlands

*Corresponding author: e.l.dejong@uva.nl

# 1. Introduction

What does understanding mean in relation to technology? Answering that question quickly leads to a nuanced position: it seems to depend on the context. A taxi driver may 'understand' their car well enough to operate it, but this understanding may prove inadequate when the car breaks down. Similarly, a shipyard owner may have a deep understanding of their vessels' capabilities but lack the understanding required to sail them. Understanding in relation to technology is thus not a singular or uniform concept; rather, it varies depending on one's role and goal in relation to a given technology.

Recognising the situated and context-dependent nature of 'technological understanding' is crucial for guiding efforts to improve people's understanding of a technology, whether it is in education, policymaking or broader public engagement with technological developments. For instance, while the emergence of second-generation quantum technologies has prompted calls to promote public understanding (Vermaas, 2017; Coenen & Grunwald, 2017: p.292; Seskir et al., 2023; Roberson, 2023; Rathenau Instituut, 2023: p.3; Nationale Agenda Quantum Technologie, 2019: p.40), what such understanding entails remains unclear. Does it refer to a simplified, 'light' version of expert understanding, or does it demand a qualitatively different kind of understanding? What does it even mean to be an expert in relation to technology?

To address these questions and develop a more precise account of understanding in relation to technology, we use De Jong and De Haro's (2025) notion of *technological understanding* and seek to specify it further. Technological understanding refers the kind of understanding that is involved in making and using technology. Drawing an analogy with scientific understanding (De Regt, 2017), De Jong and De Haro define it as the ability to (recognise how to) realise an aim by using a technological artefact[2]. Thus, technological understanding is a cognitive skill required for and demonstrated by successfully employing a technological artefact to achieve an aim.

What constitutes 'using a technological artefact'—and, consequently, when technological understanding is achieved—depends in part on the context. Different contexts impose different requirements on what counts as sufficient understanding, depending on the agent's role and goals: recall the distinction between the taxi driver and a mechanic or the shipyard owner and a captain. To advance the notion of technological understanding and assess when it has been (sufficiently) achieved, the general concept can and should be further specified for particular contexts.

De Jong and De Haro took a first step by specifying understanding within the context of technological design. In this context, the ability to successfully use a technological artefact entails the ability to design it—thinking up and constructing an artefact capable of achieving the desired aim. This paper seeks to further develop the notion of technological understanding by specifying it for two additional contexts, each representing a typical mode of engagement with technology: 'operation' (practical use) and 'innovation' (devising applications). Together, these three specifications or 'types' of understanding provide a more comprehensive account of technological understanding.

The paper is organised as follows. Section 2 summarises De Jong and De Haro's original account of technological understanding, and further develops some aspects of it: it highlights its pragmatic and gradual character, and distinguishes three main contexts in which technological understanding applies. Section 3 specifies technological understanding for these contexts, resulting in three context-specific types of understanding. To further clarify the nature of technological understanding and its specifications, Section 4 proposes an approach for assessing it based on the ability for counterfactual reasoning. Finally, Section 5 outlines the implications of this framework for discussions about the societal aspects of a technology, arguing that such discussions are epistemically conditioned and require (a specific type of) technological understanding.

[2] The term 'technological artefact' refers to a specific type of device, in contrast to 'technology' in a general sense.

# 2. The context-dependence of technological understanding

In this Section, we introduce the general concept of technological understanding (Section 2.1) and highlight its contextual, pragmatic and gradual character (Section 2.2). We then expand on this conceptual and gradual character by describing three main contexts in which technological understanding applies (Section 2.3).

## 2.1 The conception of technological understanding

According to De Jong and De Haro (2025), the ability to realise an aim by using a technological artefact involves the cognitive skill of understanding. They draw an analogy with scientific understanding: Just like scientific understanding is the ability *to use a theory to explain a phenomenon*, technological understanding is the ability *to use a technological artefact to exploit a phenomenon for a practical*[3] *aim*. In both cases, the cognitive skill of using a tool to achieve something—whether explanation or any practical end—is central.[4] By analogy with De Regt's *Criterion for Understanding Phenomena* (De Regt, 2017), they propose the *Criterion for Technological Understanding* (CTU):

**CTU:** *An aim $A$ is technologically understood if it can be realised by using a technological artefact $t$.*

Here, a technological artefact is formally defined as '*a designed object, consisting of a physical structure $X$ that produces a physical phenomenon $P$ to achieve a direct aim $a$, which in turn serves an ultimate aim $A$'* (De Jong & De Haro, 2025: p. 9). The artefact thus 'embodies' the exploitation of a physical phenomenon for a practical purpose. This conception builds on the accounts of Kroes (2002) and Houkes and Vermaas (2010), emphasising the dual nature of technological artefacts: their *physical* and *intentional* aspects. By characterising a technological artefact as 'a designed, functional object', De Jong and De Haro (p. 6) specify the artefact's physical aspect as the combination of a physical structure and a physical phenomenon, and its intentional aspect in terms of the artefact's direct and ultimate aims.

Taking the example of a laser, the operation of $t$ (i.e., the laser) fulfils a direct aim (i.e., producing a concentrated beam of light), which in turn serves an ultimate aim (e.g., the ability to cut and weld materials). The definition of technological understanding as the ability to realise an aim A by using $t$ thus means: Realising an aim $A$ by producing a phenomenon $P$ through a physical object $X$ that achieves a direct aim $a$.[5] This detailed definition is especially relevant when we talk about the required level of 'intelligibility' of the artefact.

The ability to use $t$, which is central to technological understanding, requires that the artefact be *intelligible* to the subject $S$. Thus, the criterion for technological understanding can be alternatively phrased as follows: *An aim $A$ is technologically understood if it can be realised by an* intelligible *technological artefact $t$.* De Jong and Haro explain $t$'s intelligibility as its 'useability': the combination of features of the artefact that enable the user to recognise its quality as an effective means to achieve an aim $A$. To test whether an artefact is intelligible, they propose a *Criterion for Intelligible Technological Artefacts* (CITA)[6]:

> **CITA:** *A technological artefact $t$ is intelligible for a subject $S$ (in context $C$) if they can recognise qualitatively characteristic consequences of $t$'s operation without practically performing its operation.*

[3] Following De Jong and De Haro's original account, we call the aim 'practical' since its realisation requires action.
[4] In the case of scientific understanding, the thing to be understood is the phenomenon—and not the theory itself—, whereas in the case of technological understanding, the object of understanding is the practical aim—and not the artefact itself. However, as noted in De Jong and De Haro (2025), since technologically understanding an aim requires qualitative insight into the consequences of the artefact's operation, it could also informally be explained as 'understanding a technology'.
[5] This conception of a technological artefact, and by extension, the endorsed account of technological understanding, directly applies to physical artefacts. The account can be extended to apply to software technologies as well, not least because these build on hardware technologies. Expanding the account would, however, require further theoretical elaboration, which is beyond the scope of this paper.

An important aspect of 'recognising qualitatively characteristic consequences of $t$*'s* operation' is that it entails the ability to (counterfactually) reason about $t$. Thus an artefact is intelligible if one is able to reason about the results of $t$'s operation and recognise how to use it to realise the aim $A$. This counterfactual aspect of intelligibility will be particularly useful in Section 4, when we propose an approach to assessing technological understanding.

## 2.2 Pragmatic and gradual aspects of technological understanding

Characterising understanding as a cognitive skill highlights its pragmatic nature. It is the ability to *use* knowledge (Reutelinger, Hangleiter & Hartmann, 2018; De Regt, 2017; Grimm, 2012; Wilkenfeld, 2013; Hills, 2015), and thus a cognitive achievement: a successful activity. Similarly, technological understanding emphasises the ability to achieve something: realising an aim by using a technological artefact.

There are three key implications that follow from characterising understanding as a pragmatic notion. First, as a skill, understanding always involves a subject—someone for whom the artefact is intelligible. However, we use the term 'agent' rather than 'subject' to emphasise that understanding entails the ability to do something. The implication of an agent highlights that understanding or using knowledge occurs within a specific context: their role and goal determine what constitutes understanding in that situation, making it inherently contextual. Moving beyond the individual agent (microlevel), we focus on the 'typical agent' at the mesolevel—an agent representing a class of individuals defined by shared roles and goals. These roles and goals, in turn, establish the standards for sufficient understanding and the skills required of the typical agent. We will further elaborate on this concept of 'context' and the corresponding typical agent in the next section.

Second, recognising the pragmatic nature of understanding makes it more natural to conceive of technological understanding as a *gradual* notion. As an ability, understanding can come in degrees, as also suggested by other authors (Kelp, 2015; Baumberger, 2019; De Jong & De Haro, 2025). This aligns with the intuition that technological understanding can develop over time, or that one agent may possess a more advanced understanding than another. The idea that understanding can come in degrees can be traced to the degree of intelligibility: The more intelligible the artefact is, the more (complex) actions the agent can perform using it. Thus, an agent's technological understanding directly relates to the extent to which the artefact is intelligible to them. This gradual character further strengthens the argument for technological understanding as a skill that is, in principle, trainable and testable.

This leads to a third point: technological understanding, conceived as a practical concept, allows for the assessment of its degree based on an agent's ability to perform specific tasks. By conceptualising technological understanding in terms of actions, we highlight its observable aspects, providing a foundation for task-based assessment (see Barman et al., 2024 for scientific understanding). This testable nature of technological understanding enables, in principle, the development of metrics and benchmarks. While the aim of this paper is not to develop a quantitative metric, we will discuss various levels of understanding in Section 3, and explore the possibility of assessing them in Section 4.

[6] This criterion mirrors the Criterion for Intelligible Theories (De Regt, 2017). According to this criterion, a scientific theory is intelligible for a subject S (in context C), if they can reason about the consequences of a theory without doing exact calculations.

## 2.3 Three typical contexts where technological understanding applies

As introduced in Section 2.2, the fact that technological understanding implies a 'user' necessitates consideration of the context of use, including the goals and skills of the (typical) agent. By 'context,' we refer to a specific mode of engaging with technology, encompassing practices established by a (technological) community. Rather than denoting a particular situation, a context represents a broad category of activities that require the same type of technological understanding. Moreover, whether an agent has attained (a sufficient level of) technological understanding can only be determined within a specific context. In this section, we distinguish three key contexts where technological understanding applies.

In addition to the context of design, we introduce the contexts of 'operation' and 'innovation.' These three contexts are distinguished based on three logical possibilities regarding the scope of the general (overarching) concept of technological understanding. Beyond the agent, the central elements in technological understanding are the ultimate aim $A$ and the artefact $t$. The general notion leaves open the question of whether technological understanding includes the selection of $A$ and/or $t$, i.e. whether these elements are predetermined or fixed. De Jong and De Haro (2025) outline three logical options for the scope of technological understanding:

(i) choosing neither $A$ nor $t$ (so that $A$ and $t$ are both *open*): this is the *maximal account*;
(ii) choosing both $A$ and $t$ (so that $A$ and $t$ are both *fixed*): this is the *minimal account*; and
(iii) choosing either $A$ or $t$.

We argue that these three logical options correspond to three typical contexts of engagement with technology, each representing a specific conception or type of technological understanding (*Table 1*, p.6).

In the first case, (i), neither the aim nor the artefact is given: technological understanding, in this case, includes the process of selecting a possible aim $A$ and conceiving a suitable artefact $t$ in response. This represents the maximal account, offering the greatest degree of freedom. We suggest that this variant of technological understanding is most closely associated with the context of 'technological innovation,' where new connections are created between a specific aim and a technological artefact. Activities aimed at developing innovative technological solutions to a given aim fall within this context, encompassing a wide range of pursuits, from entrepreneurial innovation to innovation policy.

In the other extreme case, (ii), both the aim and the artefact are predetermined: technological understanding does not involve the selection of either the aim $A$ or the artefact $t$. This represents the minimal account, as the scope of technological understanding is more narrowly defined. This variant typically applies to the context of practical use, where a technological artefact is employed for a common purpose. We encounter this context of practical use whenever we utilise a technological artefact to accomplish a specific task. We will refer to this context as one of 'operation.'

In the remaining case (iii), the ultimate aim $A$ is given, but the artefact $t$ is not.[7] This constitutes the *via media*—an intermediate option between the maximal and minimal accounts. In this situation—where the goal is fixed but the artefact is yet to be determined—the focus is on selecting a suitable artefact $t$ to achieve the given aim $A$. This case is typical of the design context. Aligning with De Jong and De Haro (2025), we define the context of design as encompassing all practices involved in translating a practical aim into a functional object, as well as the process of physically constructing the artefact (pp.16-17).

[7] Technically speaking, there is a second variant of the *via media,* where the artefact $t$ is fixed, but the aim $A$ remains open. We categorise this scenario under (i), as the artefact cannot truly be considered 'fixed' if the aim is unspecified. Formulating an aim inherently involves selecting the artefact—whether by choosing an existing artefact, adapting it to the given aim, or designing a new one. In all these cases, the artefact to be used is not genuinely fixed. If the aim $A$ remains open, the artefact $t$ is likewise open. Conversely, selecting an artefact does not necessarily entail the formulation of an ultimate aim.

The three logical options for the scope of technological understanding and the three typical contexts to which they correspond are summarised in *Table 1* (below).

| *Logical option* | | *Aim* | *Artefact* | *Context* |
|---|---|---|---|---|
| (i) | maximal account | open | open | **Innovation** |
| (ii) | minimal account | fixed | fixed | **Operation** |
| (iii) | via media | fixed | open | **Design** |

*Table 1. Three typical contexts in which technological understanding applies, depending on whether the aim and the artefact are open (underdetermined) or fixed (predetermined).*

## Intelligibility and degrees of understanding

Intelligibility is a prerequisite in all contexts for the successful use of the artefact $t$. Recall that for an agent to use the artefact $t$, $t$ must first be intelligible to them. However, intelligibility itself is a contextual concept: whether an artefact is deemed to be (sufficiently) intelligible depends on the specific context. Different contexts and aims impose varying demands on the intelligibility of a technological artefact.

This can be most clearly understood in terms of the components of the artefact $t$ that are emphasised by the notion of technological understanding in a given context —specifically, the physical structure $X$, the phenomenon $P$, the direct aim $a$ and the ultimate aim $A$. These components interact differently depending on the context of use, which means that that the artefact $t$ may be (primarily) intelligible in terms of $X$ and $P$, or in terms of $a$ or $A$.

This distinction pertains to the focus of counterfactual reasoning involved. Technically, the direct aim $a$ emerges from the interplay between $X$ and $P$; however, by focussing on $a$, we frame the phenomenon in terms of a specific objective. We shift the focus from, for example, the technical details of a laser's operation (i.e., stimulated emission) to its practical result (i.e., a concentrated light beam). If, given the context, the direct aim $a$ is the primary consideration for the intelligibility of $t$, then the agent will predominantly derive the consequences of the artefact's operation in terms of $a$, with less emphasis on its physical structure $X$ and the phenomenon $P$ it produces. In contrast, another context may require the artefact's intelligibility at the level of both $X$ and $P$, in which case the agent will reason about how the interaction between $X$ and $P$ can be used to achieve the ultimate aim $A$.

Within any given context, the intelligibility of a technology can be understood as existing on a continuum. This means that intelligibility comes in varying degrees, which in turn reflects different levels of technological understanding. Each type of technological understanding, therefore, is an ability that can be possessed to varying extents and should be treated as a spectrum.

The threshold for what constitutes 'sufficient' understanding is contingent upon the specific context. For instance, in the context of operation, the level of technological understanding required for professional use may differ significantly from that required in a private setting. Therefore, the criteria for defining technological understanding are inherently context-dependent, with varying thresholds depending on the role and goals of the typical agent. As we discussed earlier, specific roles and goals in engaging with technology can also vary within a given context. (For further elaboration on these roles, see [reference edited]) *Figure 1* illustrates the various contexts and their corresponding types of technological understanding, emphasising the notion that understanding exists on a spectrum.

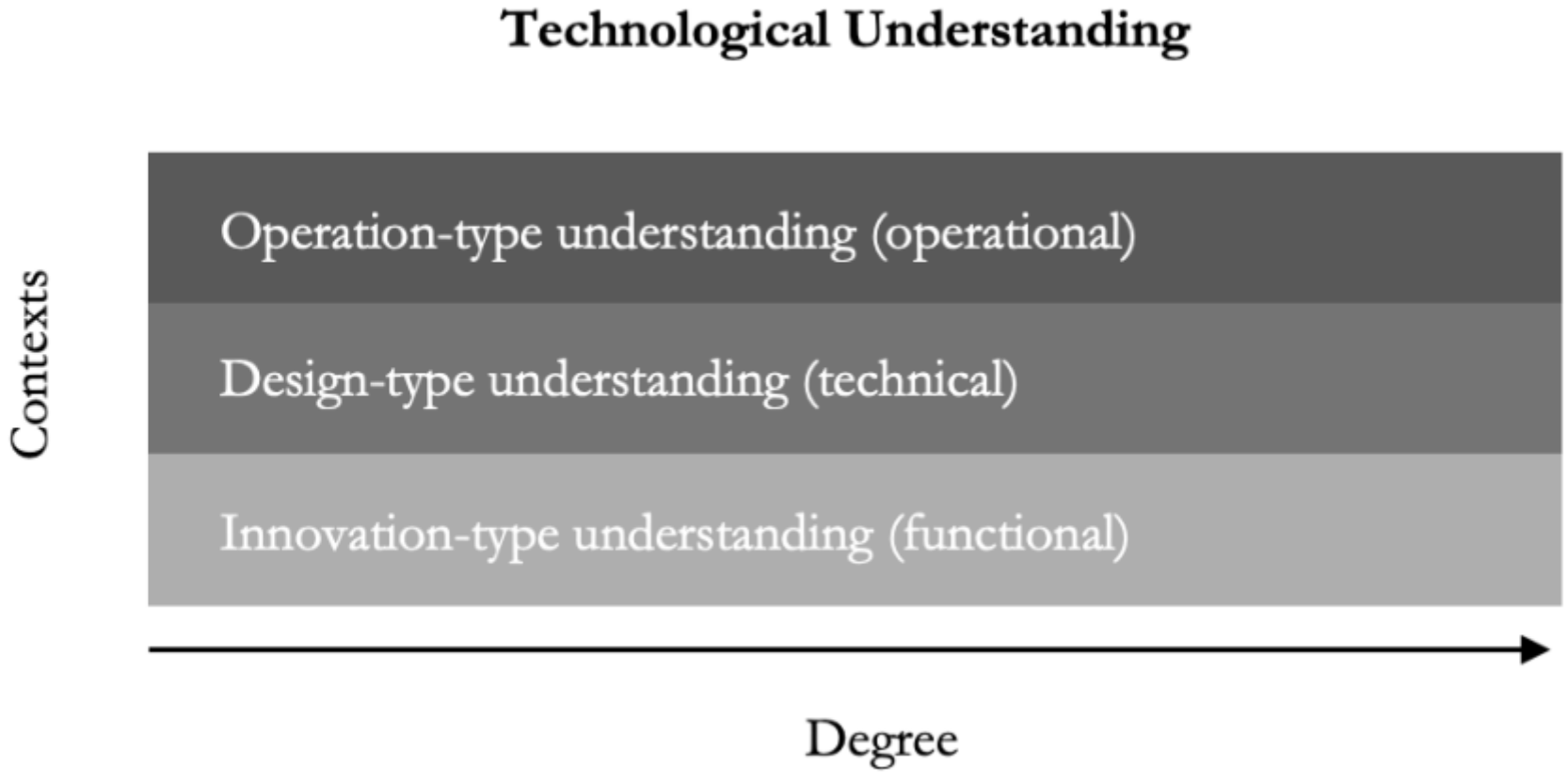

*Figure 1. Three specifications of technological understanding in three different contexts.*

### Consequences for expertise

The contextual and gradual view of technological understanding—as an ability that varies in both type and degree—has significant implications for the concept of expertise. If we conceive expertise as possessing an advanced level of understanding, then, given the context-dependence of understanding, expertise itself is also context-dependent. The idea that different types of technological understanding apply to different contexts allows for multiple interpretations of expertise, shaped by the specific context.

Beyond context, the roles and goals of an agent influence what qualifies as expertise. For instance, in an operational context, the same level of understanding might be seen as expert in one scenario but insufficient in another. Just as the threshold for 'sufficient understanding' varies across roles, so does the definition of expertise—what qualifies as 'advanced' understanding is role- and goal-dependent.

This context-dependent, agent-relative view of expertise aligns with common intuitions: a designer may be considered an expert in a laboratory setting but not necessarily in everyday operational contexts. Reexamining expertise is a key implication of distinguishing between different types of technological understanding. We will further explore these implications in Section 5.

*Key points Section 2: The context-dependence of technological understanding*

- **Technological understanding:** Technological understanding is the cognitive ability that enables the successful achievement of an aim by using a technological artefact.
- **Pragmatic, contextual and gradual notion:** This ability has a pragmatic, contextual and gradual character: it emphasises the ability to do something (rather than to know something), which is relative to a particular context and can come in degrees.
- **Three typical contexts:** Based on whether the aim and the artefact are predetermined or underdetermined, three typical contexts can be distinguished in which technological understanding applies: a context of design, of operation, and of innovation.

# 3. Specifying technological understanding: three types

Drawing on the distinction between the three contexts of technological understanding, this section further elaborates on the three corresponding types of understanding. We first summarise the design-type of understanding ('technical understanding') (3.1) and then develop the operation-type ('operational understanding) (3.2) and the innovation-type ('functional understanding') (3.3).

## 3.1 Design-type technological understanding

We first consider the design context, pre-eminently corresponding to the option whereby the aim is given and the artefact is unspecified (option (iii), *via media*). Technological understanding requires the availability of an artefact that can realise that aim. This is, however, not a given. Often there is only an aim and the desire for an artefact that can achieve it. In this case, 'using an appropriate artefact' requires *designing* one (De Jong & De Haro, 2025).

More formally, successfully designing an artefact $t$ involves the ability to construct a physical structure $X$ that produces a phenomenon $P$ that achieves a direct aim $a$, which realises the ultimate aim $A$. In a design context, technological understanding thus manifests as the ability to develop an artefact suited to a given practical aim. Conversely, the ability to design an appropriate artefact both requires and reflects technological understanding.

The type of technological understanding involved in the design context is defined as follows:

> **Technological understanding(design):** *The ability to realise an aim* $A$ *by* **designing** a technological artefact $t$.

### Intelligibility of the technological artefact

Achieving technological understanding requires the artefact to be intelligible. In a design context, this means the ability to reason about the artefact at a technical level. A designer must recognise how an artefact's consequences stem from its design. This design-type of understanding can be referred to as *technical understanding*.

From the perspective of technological understanding's key components, technical intelligibility emphasises the ability to derive the artefact $t$'s consequences in terms of its structure $X$ and the active phenomenon $P$. In other words, if intelligibility concerns knowing how $t$ can be used to achieve $A$, then in a design context, it means understanding how the specific relationship between $X$ and $P$ contributes to $A$.

Since intelligibility exists on a spectrum, different levels of design-type technological understanding can be distinguished. The required depth of intelligibility depends on an agent's role and goals. For example, for an electrical engineer designing a technological artefact (e.g., a smartphone or a lift), the artefact needs to be intelligible to a more advanced extent than for a technician responsible for its implementation or repair.[8]

## 3.2 Operation-type technological understanding

The second context we identified is that of practical use, or 'operation', corresponding to the option where both the aim and the technological artefact are relatively fixed (option (ii), minimal account). By 'practical use,' we refer to the direct operation of a technological artefact with the intention of achieving a specific practical aim. Here, 'using an artefact' is specified as operating it. This leads to our second specification of technological understanding within the operational context:

> **Technological understanding(operation):** *The ability to realise an aim $A$ by* ***operating*** *a technological artefact $t$.*

[8] It is true that practical implementation of technology requires additional knowledge and skills pertaining to specific contextual factors, but the required depth of knowledge about the physical structure and the phenomena that compose a given artefact is less than for design practices.

It is important to highlight from the outset that the operational context is highly diverse. It encompasses everything from the professional use of complex machines—such as MRI scanners, industrial lasers, or cranes—to the everyday operation of simpler devices by consumers, such as electric toothbrushes, computers, or cars. However, as we discussed in Section 2.3, a unifying perspective emerges when we focus on the type of technological understanding involved in these cases. However, as we discussed in Section 2.3, a unifying perspective emerges when we consider the type of technological understanding involved. Despite the variety of technologies and their uses, all forms of operation rely on a specific kind of technological understanding. While such differences may be relevant, for example, in the context of setting up licensing for use, this paper will not distinguish between different types of operation, though we acknowledge that such distinctions exist. Instead, we will primarily consider these differences in terms of varying degrees of the operation-type ('operational') understanding.

We should also emphasise that operating an artefact typically requires additional practical skills, tailored to the specific circumstances that affect the artefact's operation. Technological understanding encompasses the epistemic dimension of the activity: the cognitive ability to recognise how an artefact can be operated, rather than the physical ability to do so.

For example, operating a laser involves understanding how the laser can be used to cut material. While this cognitive understanding is necessary, it is not always sufficient for successful operation, as physical handling skills are also required. Therefore, successful operation demonstrates both cognitive understanding and physical skills, with the latter not being part of the cognitive skill itself.

### Intelligibility of the technological artefact

In the operation context, an artefact $t$ is intelligible for a user if they can reason about it at an *operational* level—enabling them to practically use the artefact. For an artefact to be operationally intelligible, an agent must primarily recognise its consequences in terms of how it achieves the direct aim $a$, and ultimately, the broader aim $A$. The artefact only needs to be 'technically' intelligible (in terms of $X$ and $P$) to the extent that this understanding is necessary for its successful operation.

This level of practical reasoning differs from the technical reasoning required by designers. For practical use, extensive knowledge of the physical phenomena underlying an artefact's operation is often not necessary. For instance, one can send an email without understanding how a CPU or the internet works. However, in some cases, reasoning about an artefact and its operational consequences does require some insight into its internal mechanisms. Operating a crane necessitates an understanding of gravitational forces. The required depth or quality[9] of this insight depends on the specific role and goals of the agent. Ultimately, the user needs to understand enough about the inner workings of the artefact to recognise how it can be used successfully.[10]

---

[9] Is it, for example, enough to *know that* something is the case (e.g., knowing *that* gasoline will destruct my diesel engine) or is it required to understand why something occurs? For technological understanding, knowing-that is not enough: one should be able to see why something is the case. However, the range of why-questions that one should be able to answer, depends on the circumstances. We expand on this idea of technological understanding as the ability to answer why-questions in Section 4.

[10] We adopt the view that, generally, technological understanding requires scientific understanding (for a more detailed discussion, see De Jong & De Haro, 2025). Both notions come in degrees and have contextual standards, so the depth of the required scientific understanding is attuned to the type and degree of technological understanding.

Operational intelligibility should also be understood as a matter of degree: for an advanced user, the artefact will be more intelligible than for an inexperienced one, resulting in a higher or lower level of operational understanding. The required level of understanding depends on the agent's role, goals, and the specific circumstances. For example, driving a car on flat tarmac may require less technological understanding than driving off-road. Similarly, using a computer in a professional context generally demands a higher level of proficiency than using it for personal tasks.

## 3.3 Innovation-type technological understanding

The third context we distinguished is the context of innovation, where both the ultimate aim $A$ and the artefact $t$ are open (option (i), maximal account). While the designer's role is to identify an artefact that can achieve a given aim, and the end-user operates the artefact to achieve a specific aim, the innovator defines the aim to be achieved and conceptualises the technological response by selecting (or adapting) an existing artefact or conceiving a new one yet to be designed.

This requires a deeper, 'embedded view' of the ultimate aim than is needed in design or operation contexts. The innovator must possess a comprehensive understanding of the aim's nature and its societal implications, enabling them to assess whether, and to what extent, a technological artefact can fulfil it.

While the concept of innovation lacks a universally accepted definition (Bontems, 2014), it is generally associated with novelty and challenging the status quo. The Cambridge Dictionary defines innovation as "the creating and use of new ideas or methods," while the Oxford Dictionary describes it as "the process through which new products, concepts, services, methods, or techniques are developed." Furthermore, innovation is frequently linked to the creation and application of technology (Bontems, 2014; Blok & Lemmens, 2015; OECD, 2005).

Although a comprehensive discussion of innovation is beyond the scope of this paper (for a philosophical exploration of the concept, see, for example, Blok, 2021), we adopt a broad conception of innovation as the ideation of a technological response to a desire to change the status quo. This definition views innovation as an activity rather than a specific invention, focussing primarily on the ideation of opportunities for the successful application of a technological artefact rather than its actual creation.

This conception of innovation incorporates elements of both the design context and the context of practical use. Like a designer, an innovator establishes a connection between an aim and an artefact that did not previously exist. However, unlike the designer, the innovator does not directly create or design the artefact. Instead, the innovator "designs" the relationship between the artefact and the aim. This involves conceptualising, outlining, or describing in broad strokes a potential relationship. In essence, the innovator does not design the technological artefact itself but focusses on devising how a (yet to be designed) artefact might be applied to achieve a specific aim.

Like the end-user, the innovator is primarily focussed on the ultimate aim $A$ and considers the artefact $t$ as a potential means to that end. Successfully doing so requires not only insight into the aim but also an understanding of the sociotechnical context in which the artefact must function (De Jong, 2022: pp.14-16). However, unlike the end-user, the innovator is not necessarily the one who will practically use the artefact and does not follow a pre-existing $t$-for-$A$ relationship. In fact, the innovator determines the problem, need, or demand and conceives a technological response to address it.

Thus, the innovator 'devises' the artefact as a potential means to an aim, with 'devising' referring to the process of envisioning or ideating a technical response (e.g., the use of MRI) to address a specific need (e.g., non-invasive examination of soft body tissues). This means they must be able to align the artefact's capabilities with an aim and assess its performance in relation to that aim. The innovation-type of technological understanding is therefore capability-oriented, focussing on the functionalities and potential applications of the artefact. We can thus refer to this type of understanding as 'functional understanding.'

Innovators may conceive entirely new technologies, such as generative Artificial Intelligence and quantum computers, or discover novel applications for existing artefacts. For instance, radar communication systems were repurposed to heat food, leading to the invention of the microwave. Similarly, Magnetic Resonance Imaging (MRI), originally designed for medical diagnostics, has been adapted for food quality assessment. Also, before the combustion engine was invented, innovators attempted to repurpose railway steam engines for early steam-powered cars. Innovators play a key role in (re)purposing and (re)imagining technological artefacts, extending them into new domains and applications.

Considering the innovator as a typical agent provides us with a third specification of technological understanding, specifically applicable to an innovation context:

> **Technological understanding(innovation):** *The ability to realise an aim* $A$ *by* ***devising*** *a technological artefact* $t$.

## Intelligibility of the technological artefact

For an innovator to devise an artefact and connect its capabilities to a specific aim, the artefact must be intelligible at an appropriate level. In an innovation context, intelligibility entails the ability to reason about the artefact in an evaluative manner: like the operator, the innovator must recognise the artefact's consequences, particularly in terms of how effectively the achievement of a direct aim $a$ contributes to an ultimate aim $A$. However, the innovator's required insight into the problem, need, or demand that the artefact is meant to address is necessarily deeper.

At this level, the artefact must be intelligible in terms of its functional properties rather than its inner workings or immediate, concrete effects. To invent a use case, the innovator must grasp the artefact's capabilities and potential applications. This aligns with Houkes and Vermaas's (2010) function analysis, which defines ascribing a function to an artefact as assigning to it a specific (physicochemical) capacity (p. 84). Functional intelligibility can therefore be understood as *epistemic access to the artefact's capabilities*[11]. In some cases, depending on the agent's role and goal, devising a technological artefact may also require insight into how these capabilities emerge from its design and use.

The degree of functional intelligibility required varies according to the agent's role, goals, and circumstances. A strategic visionary—who conceives a novel relationship between an artefact and an aim—may require a different depth of understanding than a normative assessor, who evaluates the artefact's use, not only in relation to its intended aim but also within a broader societal or ethical framework. Each of these roles demands a tailored level of intelligibility and, consequently, a corresponding degree of understanding.

Now that we have outlined three types of technological understanding, it is important to emphasise that they are not mutually exclusive; in practice, they often overlap or intertwine. A user may demonstrate innovation by employing an artefact in unforeseen ways, an innovator may engage in design by co-formulating design requirements, and a designer may enter the sphere of innovation by (co)determining the ultimate aim. The conceptual distinction between these types of understanding primarily serves to highlight which form of understanding is most prominent in a given context.

Beyond clarifying these distinctions, this framework also has normative significance. It provides a basis for arguing that certain types of technological understanding are necessary to undertake specific actions within each context. Moreover, within any given context, the degree of technological understanding required depends on the agent's role and goals. This perspective—that different types and degrees of understanding are required for different actions—underscores the need for a method to assess technological understanding. In the next section, we propose a possible approach to this assessment.

[11] We favour the use of the term 'capabilities' rather than 'capacities', since 'capabilities' refers to the qualitative aspects of a technology's potential abilities and functionalities, whereas 'capacities' is often associated with the quantitative limits or extents of these abilities.

*Key points Section 3: Three context-specific notions of technological understanding*

- **Design context:** An agent has design-type or 'technical' understanding if they can *design* a technological artefact $t$ to realise an aim $A$.
- **Operation context:** An agent has operation-type or 'operational' understanding if they can *operate* a technological artefact $t$ to realise an aim $A$.
- **Innovation context:** An agent has innovation-type or 'functional' understanding if they can identify an aim $A$ and *devise* a technological artefact $t$ to realise it.

# 4. Assessing technological understanding

To clarify the nature of the different types of technological understanding, we propose an approach for assessing it through counterfactual reasoning (4.1), building on a similar benchmark for scientific understanding (Barman et al., 2024), and argue that different types of technological understanding correspond to distinct sets of relevant counterfactuals (4.2).

## 4.1 Technological understanding as a form of inferential reasoning

The pragmatic nature of understanding—as the cognitive skill to use knowledge— suggests that it can be assessed through an agent's ability to perform specific tasks. Technological understanding is evidently demonstrated by the successful use of a technological artefact. However, to further illuminate the cognitive dimension of this achievement, we aim to make explicit the underlying cognitive tasks involved in this process. To do so, we once again draw an analogy with scientific understanding.

The argument for scientific understanding is as follows:

1a) **Proper use of a theory as the core ability:** In the case of scientific understanding, the fundamental cognitive task is knowing when and how to apply a scientific theory to explain a phenomenon. More specifically, scientific understanding involves recognising the conditions under which a given theory is relevant and how it can be used to generate explanations. For example, understanding the motion of an object under different forces means knowing when and how to apply Newton's laws to predict its motion under various conditions.

1b) **Proper theory use requires inferential reasoning:** The ability to properly use a scientific explanation depends on the capacity to draw counterfactual inferences (Woodward, 2003; Woodward & Hitchcock, 2003; Weslake, 2010; Grimm, 2012). Building on this, Barman et al. (2024) argue that scientific understanding[12] can be assessed through an agent's 'inferential ability'—that is, the ability to infer how the phenomenon $P$ would behave according to a theory $T$ given specific circumstances. Thus, scientific understanding, understood as the ability to use a scientific explanation, presupposes and can therefore be tested by an agent's ability to reason counterfactually.[13]

1c) **Inferential reasoning can be tested with what-if questions:** Barman et al. (2024) argue that the inferential ability central to scientific understanding can be evaluated by an agent's capacity to answer a specific type of question—namely, questions about what would have happened if certain factors were different or what would happen if specific changes were made. Successfully answering such 'what-if-things-had-been-different' and 'what-would-happen-if' questions is a key indicator of scientific understanding.[14,15] This ability involves "postulating hypothetical scenarios about what would occur under a specific set of circumstances" (Barman et al. 2024: p.5).

[12] Based on the account by De Regt (2017).

[13] While Barman et al. (2024) primarily focus on correctly applying a theory under different conditions, counterfactual reasoning also plays a crucial role in evaluating how modifications to the theory would impact its explanatory or predictive power (Woodward, 2003). In other words, counterfactual reasoning is not only essential for applying scientific theories but also for their refinement and adjustment.

Their account builds on Kuorikoski and Ylikoski (2015), who argue that "correct what-if inferences provide a natural measure of understanding" and suggest that the breadth and precision of an agent's counterfactual inferences correlate with their degree of understanding: the greater the range and accuracy of such inferences, the deeper the level of understanding.

By analogy, we propose the following argument for the cognitive skill constituting technological understanding:

2a) **Proper use of an artefact as the core ability:** Just as scientific understanding is centred on the proper use of a theory, technological understanding fundamentally entails the proper use of an (adequate) technological artefact—knowing when and how to use it to achieve an intended aim.

2b) **Proper artefact use requires inferential reasoning:** Given the structural parallels between scientific and technological understanding, the cognitive skill involved in using a technological artefact can also be understood as an inferential ability—specifically, the ability to draw inferences about the artefact's application. This means being able to predict how the relationship between the direct aim $a$ and the ultimate aim $A$ if the artefact $t$ were to be used in specific circumstances.

This involves reasoning about when and how to apply the artefact under changing circumstances, and how changes to the artefact itself could affect its performance.[16] This dual perspective is crucial for improving artefacts to better achieve their intended aims—analogous to refining a scientific theory to enhance its predictive power.

For example, successfully using a microwave to heat food requires understanding how to use it under different conditions (e.g. adjusting power levels for different types of food) and possibly how modifying the microwave (e.g. redesigning its heating elements) would impact its performance.[17]

2c) **Inferential reasoning can be tested with what-if questions**: The ability to make what-if inferences—both regarding changing circumstances and modifications to an artefact's design—is a key indicator of technological understanding and can be assessed through what-if questions. Just as what-if questions about a phenomenon $P$ test scientific understanding by gauging an agent's ability to explain $P$ under different conditions using a theory $T$, we argue that technological understanding can be tested based on an agent's ability to realise an ultimate aim $A$ under varying circumstances using an artefact $t$.

[14] More precisely, Barman et al. argue that the degree of an agent's scientific understanding of a phenomenon $P$ can be determined by assessing the extent to which the agent '(i) has a sufficient complete representation of $P$; (ii) can generate internally consistent and empirically adequate explanations of $P$; (iii) can establish a broad range of relevant, correct counterfactual inferences regarding $P$' (2024: p. 4). *Together* these abilities constitute scientific understanding—they do not represent subsequent levels of understanding. Barman et al. propose to measure these abilities through answering what-, why-, and what-if-things-had-been-different-questions. Since the ability to answer what-if-questions builds on—and therefore, includes—the ability to answer what- and why-questions, we will focus on the former as the primary indicator for understanding. This aligns with what Barman et al. refer to when they state that "the range of counterfactual inferences an agent can articulate is strongly related to the level of understanding.'"(p.5)

[15] For convenience, we will further refer to 'what-if-things-had-been-different-questions' and the related 'what-would-happen-if-questions' simply as 'what-if-questions'.

[16] In Section 4.2, it is explained that the inferential ability that focusses on changes in the artefact particularly pertains to the design-type of technological understanding.

[17] This typically applies to the context of design and possibly to the context of innovation.

These include questions about how an aim can be achieved using an artefact under different conditions—for example, how effectively a microwave ($t$) heats food ($A$) when heating a fresh versus a frozen meal—as well as questions about how changes to the artefact or its use affect its performance. For instance, how adjusting the heating time, altering power output, or redesigning the microwave's internal structure influences its ability to heat food.

The ability to infer what would happen if certain variables were changed is directly tied to the ability to bring about a desired outcome. This is why what-if questions can often be reformulated as *how* questions: *How can $t$ be used to achieve $A$ under a specific set of conditions?* This manipulative aspect of inferential reasoning (Woodward, 2003; Kuorikoski & Ylikoski, 2015) is fundamental to technological understanding, which is inherently manipulative, geared towards bringing about a certain state of affairs. In this sense, technological understanding is not merely indicated by correct what-if inferences—it *consists* in them.

The breadth and depth of an agent's technological understanding is indicated the number and complexity[18] of what-if questions they can answer correctly. The greater their ability to establish counterfactual inferences, the higher their degree of technological understanding. However, as Barman et al. (2024) acknowledge, the specific questions deemed relevant for assessing understanding are context dependent. Similarly, determining a *sufficient* level of understanding varies across different situations, depending on the demands of the task at hand.

To sum up, technological understanding, like scientific understanding, involves inferential reasoning and can be assessed by an agent's ability to answer what-if questions about the use and modification of technological artefacts under varying conditions. Since technological understanding is defined by the ability to answer what-if questions, its context-dependent nature implies that different contexts require different sets of relevant questions. We explore this further in the next section.

## 4.2 Each context of understanding requires specific inferential abilities

In the previous section, we argued that technological understanding can be assessed through the ability to answer what-if questions, which involves generating scenarios and reasoning about how an aim can be achieved by using an artefact under varying circumstances. Since each type of technological understanding corresponds to distinct abilities—namely: successfully designing, operating, or devising an artefact—different counterfactuals are relevant in each case. In this section, we will specify the set of what-if questions pertinent to each type of technological understanding.

### Assessing operation-type technological understanding

Successfully operating an artefact requires understanding how to use the artefact under varying conditions, specifically how changes in those conditions impact its performance and the achievement of the intended aim. This operation-relevant inferential ability can be assessed through what-if questions that explore how to achieve an aim under different circumstances by operating the artefact in specific ways. This includes questions like: *"What if the operating conditions changed? How would this affect both the performance and the proper operation of the artefact?"* and *"What would happen if the artefact malfunctioned?"* For example, *How different weather conditions affect my car's performance, and how could I adjust my driving to stop at the traffic sign in time?*

Answering such questions requires operational intelligibility: the artefact needs to be intelligible at the level of its direct practical results. The artefact is intelligible at this level if the operator can recognise its consequences in terms of the direct aim it realises (e.g., pressing the brake pedal results in slowing the car). However, answering operation-relevant what-if questions may also require some degree of technical intelligibility of the artefact. For example, to operate a crane properly, a crane driver needs not only to understand how weather conditions affect the lifting of an object but also how the angle of the slings influences the maximum weight that can be safely lifted.

[18] Complexity here refers to the generality of a question, where more general questions are associated with lower levels of complexity and more specific questions with higher levels.

### Assessing design-type technological understanding

Successfully designing an artefact requires understanding how changes to the artefact's structure or components affect its performance. This design-relevant inferential ability can be assessed through what-if questions that explore how an aim can be realised under specific circumstances by designing the artefact in particular ways. In a design context, the relevant what-if questions focus on how the artefact's design can achieve a given aim under varying conditions: *"What would happen if the artefact were designed differently (e.g., by using different materials)? How would this affect its performance in relation to the aim?"* For example: *What would happen if we replaced the red laser in this machine with a blue laser (shorter wavelength)? How would this affect the machine's measuring accuracy?*

Answering such questions requires *technical intelligibility*: the artefact must be intelligible at the level of its internal structure and functioning. As we discussed in Section 3, an artefact is intelligible at this level if the agent can recognise the consequences of its operation based on its physical structure $X$ and the active phenomenon $P$. Design-relevant what-if questions thus focus on deriving the artefact's consequences based on specific configurations of $X$ and $P$.

### Assessing innovation-type technological understanding

Successfully devising a technological artefact involves linking the capabilities of technologies to practical aims. This requires understanding how and to what extent an artefact's potential capabilities can realise a specific goal. In this context, relevant inferences involve reasoning about how new or existing technologies can be applied to solve problems or achieve other objectives. This includes questions like: *"What would happen if we changed the aim—how would that affect the appropriateness of the artefact as a technological response?"* and *"What would happen if we changed the artefact (partly or entirely)—how would that affect the achievement of the aim?"*

For example: *What if we integrated AI-driven diagnostics into healthcare services? How would the capabilities of AI enhance the effectiveness and reach of these services? Or what if we applied autonomous vehicle technology to public transportation systems in congested cities? How would the features of autonomous vehicles contribute to the aim of reducing traffic and improving commute times?*

Answering such questions requires functional intelligibility: the agent must recognise the artefact's capabilities and assess their relevance to specific goals ('fitness for purpose').

In an innovation context, the relevant set of what-if questions may also address the design and operation of the artefact if such questions help determine whether the artefact is an appropriate technological response to the aim. Like in the operational context, the focus is primarily on the relationship between the aim and the artefact, though in some cases, reasoning about the artefact's consequences at the level of its inner workings may be necessary for successful devising.

*Table 2* (p.16) summarises the core abilities for each of the three types of technological understanding (row 1), the corresponding level of intelligibility (row 2), the counterfactual reasoning indicative of each type (row 3), and paradigmatic what-if questions (row 4).

| | *Type of understanding* | *Operation-type (operational understanding)* | *Design-type (technical understanding)* | *Innovation-type (functional understanding)* |
|---|---|---|---|---|
| 1 | Technological understanding is the ability to realise an aim by… | …**operating** a technological artefact | …**designing** a technological artefact | …**devising** a technological artefact |
| 2 | Requires the artefact to be intelligible at the level of its… | practical results | inner workings | functional capabilities |
| 3 | Assessable by testing the inferential ability to recognise how… | …different circumstances affect the operation of the artefact | …changes in the internal structure of the artefact affect its performance | …appropriate and effective the artefact is with respect to a specific aim |
| 4 | What-if questions that can test this ability | *What if the operating conditions changed? How would this affect the performance as well as the proper operation of the artefact?* | *What would happen if the artefact was designed differently—how would this affect its performance with respect to the desired aim?* | *What would happen if we changed the aim or the artefact—how would that affect the appropriateness of the artefact as a technological response?* |

*Table 2. Three types of technological understanding, their associated level of the artefact's intelligibility, and the inferential abilities by which they can be assessed.*

*Key points Section 4: Assessing technological understanding*

- **General approach:** Technological understanding can be assessed by testing the ability to make counterfactual inferences, which in turn can be tested by answering *what-if-things-had-been-different-*questions about the successful use of a technological artefact.
- **Design-type:** In a design context, technological understanding can be assessed by testing an agent's ability to reason about how changes in the internal structure of the artefact affect its performance.
- **Operation-type:** In a context of operation, technological understanding can be assessed by testing an agent's ability to reason about how changes in external circumstances affect the successful performance and operation of the artefact.
- **Innovation-type:** In a context of innovation, technological understanding can be assessed by testing an agent's ability to reason about the appropriateness of the artefact regarding a specific aim ('fitness for purpose').

# 5. Technological understanding as foundational competence for discussions about societal impact of technology

What are possible implications of this framework, for example for (academic and public) discussions about the impacts of a technology within society? As noted in the introduction, it has been previously suggested that considering the societal impact of a particular technology requires a certain level of understanding (Vermaas, 2017; Seskir et al., 2023). However, it remains unclear what this understanding entails. We argue that the concept of technological understanding offers insight into this issue.

Debates about the societal impact of technology typically centre on the consequences of using a technology (Swierstra & Rip, 2007: p.11). We argue that this indicates that some form of technological understanding is involved, as the cognitive skill enabling one to foresee the consequences of using an artefact. The next question, then, is which type of understanding is most relevant.

In assessing the potential (positive or negative) consequences of a technology, it is essential to understand the purposes for which it may be used. This connects closely with the reasoning level of innovation-type understanding. As previously discussed, this type of understanding involves recognising the functional capabilities of an artefact—essentially, understanding what a technology can do and evaluating its suitability for specific aims. Recognising these potential applications is crucial for ethical discussions about the desirability and consequences of a technology, serving as a starting point for considering second-order consequences that could emerge from these applications.

It is important to emphasise that focussing on the functional capabilities of a technology does not imply a form of technological determinism, where our role is limited to merely 'selecting' and directing its development; rather, it allows space for considering the desirability of a technology itself, rather than assuming its inevitability. Rather, this understanding should empower individuals to shape and guide technological development in ways that align with ethical and democratic values. For, as we have seen, technological understanding underscores the importance of the various ways to relate to an aim, including the ability to choose the aim.

When considering the societal implications of a technology, innovation-type understanding plays a pivotal role: it enables us to think about possible applications and, consequently, potential impacts. More specifically, such understanding provides a foundation for ethical discussions about technology by by enabling projections of its possible uses and consequences ([reference edited). This perspective of understanding as a foundational competence for studying the societal—and ethical—aspects of technology has at least two significant implications for the epistemic nature of discussions surrounding a technology's potential impact.

Firstly, it suggests that these discussions require more than just knowledge—they demand a specific form of understanding. What is needed is the ability to envision how a technology could be applied. This cognitive skill goes beyond the mere possession of knowledge. Simply transferring factual knowledge about a technology's potential uses is in general not sufficient to enable effective participation in discussions about its societal impact; individuals must be able to imagine and anticipate its possible applications.

Secondly, recognising that innovation-type understanding plays a key role in discussions about impact helps guide efforts to learn about new technologies. To engage in these discussions, it is not always (and often not) necessary to understand the technology at the level of its internal workings or operational use. Instead, what matters most is developing an understanding of its functional capabilities.

While assessing a technology's societal impact involves a form of technological understanding similar to that required for innovation and devising a technological solution to a specific problem or need, the focus of these understandings diverges. The innovator's role typically centres on identifying potential use cases, whereas the impact assessor's role is primarily focussed on evaluating these applications rather than creating them. However, when anticipating the impact of a technology, the ability to envision potential use cases becomes more relevant, aligning more closely with the visionary capacities of the innovator. Moreover, assessing the appropriateness of a technological response often extends beyond strategic and commercial considerations to encompass broader ethical and societal implications.

Additionally, the idea that understanding the functional capabilities of a technology is sufficient for meaningful discussion may be criticised by arguing that a deeper, more technical understanding is necessary for informed debate. For instance, without understanding the underlying mechanics of a technology, participants may overlook crucial details—such as potential risks or limitations—that only become apparent at a deeper technical level. While this is a valid point, and ideally, impact discussions based on innovation-type understanding should be complemented by analyses informed by other types of understanding, a deeper, design- or operation-specific understanding is not always accessible to those without the relevant technical background. For this group, an innovation-type understanding remains essential. Furthermore, acknowledging that innovation-type understanding plays a crucial role in facilitating impact discussions can make these discussions more accessible to a broader audience.

# Conclusion

What counts as (sufficient) technological understanding depends on the context, and the role and goal of the agent. In this paper, we further developed the notion of technological understanding (De Jong & De Haro, 2025) as an ability that varies by context and degrees. We complemented its specification for a design context—as the ability to design a technological artefact—by introducing two additional contexts: operation and innovation, respectively emphasising the ability to practically use a technological artefact and to devise its applications.

Each resulting type of technological understanding requires that the artefact is intelligible to the agent, albeit at different levels. The design-type requires an understanding of the artefact's inner workings, the operation-type focusses on its practical results, and the innovation-type centres on its functional capabilities.

To further clarify the ability encompassed by (the different types of) technological understanding, we proposed an assessment approach. Building on Barman et al. (2024), who proposed a framework for measuring scientific understanding, we argued that technological understanding consists in and can be tested through the competency to establish counterfactual inferences.

Given the three types of technological understanding, relevant counterfactual reasoning differs across contexts. We suggested that design-type understanding can be assessed through what-if questions about changes in an artefact's internal structure, operation-type through what-if questions about how varying circumstances affect its performance, and innovation-type through questions about an artefact's effectiveness and appropriateness for a specific aim. This framework provides not only a method for assessing technological understanding but also a foundation for articulating required degrees of understanding in specific contexts.

Although the distinctions between 'technical' (design-type), 'operational' (operation-type), and 'functional' (innovation-type) understanding may seem intuitive, it is important to explicitly distinguish them, since they are easily conflated in practice. Such conflations can misguide efforts aimed at improving technological understanding and wrongly (dis)qualify certain forms of understanding. Conceptually distinguishing them—while acknowledging their possible overlap— sharpens our focus on the required type of understanding, for example when discussing a technology's societal impact. Moreover, explicitly recognising different types of technological understanding fosters a more pluralistic perspective on expertise in relation to technology, acknowledging multiple types of expertise, beyond technical expertise. Technological understanding is thus not a one-size-fits-all skill; it is a contextual competence—one that is essential for any successful form of engagement with technology.

## References

Barman, K. G., Caron, S., Claassen, T., & De Regt, H. (2024). Towards a benchmark for scientific understanding in humans and machines. *Minds and Machines*, *34*(1), 1-16.

Baumberger, C. (2019). Explicating objectual understanding: Taking degrees seriously. Journal for General Philosophy of Science, 50(3), 367–388.

Blok, V. (2021). What is innovation? Laying the ground for a philosophy of innovation. *Techne: research in philosophy and technology*, *25*(1), 72-96.

Blok, V., & Lemmens, P. (2015). The emerging concept of responsible innovation. Three reasons why it is questionable and calls for a radical transformation of the concept of innovation. *Responsible innovation 2: Concepts, approaches, and applications*, 19-35.

Bontems, V. K. (2014). What does Innovation stand for? Review of a watchword in research policies 1. *Journal of Innovation Economics & Management*, (3), 39-57.

Coenen, C., & Grunwald, A. (2017). Responsible research and innovation (RRI) in quantum technology. *Ethics and Information Technology*, *19*, 277-294.

De Jong, E. (2022). Own the unknown: an anticipatory approach to prepare society for the quantum age. *Digital Society*, *1*(2), 15.

De Jong, E., & De Haro, S. (2025). Technological understanding: On the epistemic dimension of designing technological artefacts. arXiv:2503.01617.

De Regt, H. W. (2017). *Understanding scientific understanding.* Oxford University Press.

Grimm, S. (2012). The value of understanding. *Philosophy Compass*, *7*(2), 103-117.

Hills, A. (2016). Understanding why. *Noûs*, *50*(4), 661-688.

Houkes, W., & Vermaas, P. E. (2010). *Technical functions: On the use and design of artefacts* (Vol. 1). Springer Science & Business Media.

Kelp, C. (2015). Understanding phenomena. *Synthese*, *192*(12), 3799-3816.

Kroes, P. (2002). Design methodology and the nature of technical artefacts. *Design Studies* 23(3), pp. 287–302.

Kuorikoski, J., & Ylikoski, P. (2015). External representations and scientific understanding. *Synthese*, *192*, 3817-3837.

Nationale Agenda Quantum Technologie (2019). *Quantum Delta NL*. Retrieved from https://www.rijksoverheid.nl/documenten/brochures/2020/02/17/nationale-agenda-quantumtechnologie

OECD (2005). *Oslo Manual: guidelines for collecting and interpreting innovation data*. Third edition. OECD & Statistical Office of the European Communities.

Rathenau Instituut (2023). *Quantumtechnologie in de samenleving*. Rathenau Scan. Rathenau Instituut: Den Haag. Retrieved from https://www.rathenau.nl/nl/digitalisering/rathenau-scan-quantumtechnologie-de-samenleving

Reutlinger, A., Hangleiter, D., & Hartmann, S. (2018). Understanding (with) toy models. *The British Journal for the Philosophy of Science*, *69*(4), 1069-1099.

Roberson, T. (2023). Talking about responsible quantum:"Awareness is the absolute minimum that… we need to do". *NanoEthics*, *17*(1), 2.

Seskir, Z. C., Umbrello, S., Coenen, C., & Vermaas, P. E. (2023). Democratization of quantum technologies. *Quantum Science and Technology*, *8*(2), 024005.

Swierstra, T., & Rip, A. (2007). Nano-ethics as NEST-ethics: patterns of moral argumentation about new and emerging science and technology. *Nanoethics*, *1*, 3-20.

Vermaas, P. E. (2017). The societal impact of the emerging quantum technologies: a renewed urgency to make quantum theory understandable. *Ethics and Information Technology*, *19*, 241-246.

Weslake, B. (2010). Explanatory depth. *Philosophy of Science*, *77*(2), 273-294.

Wilkenfeld, D. A. (2013). Understanding as representation manipulability. *Synthese*, *190*, 997-1016.

Woodward, J., & Hitchcock, C. (2003). Explanatory generalizations, part I: A counterfactual account. *Noûs*, *37*(1), 1-24.

Woodward, J. (2005). *Making things happen: A theory of causal explanation*. Oxford university press.